\documentclass[runningheads,orivec]{llncs}
\usepackage[T1]{fontenc}
\usepackage{graphicx}
\usepackage{amsmath}
\graphicspath{{figures}}
\usepackage{hyperref}
\usepackage{color}

\usepackage[utf8]{inputenc}
\usepackage{textgreek}
\usepackage{multirow}

\usepackage[normalem]{ulem}
\usepackage{soul} 
\usepackage{xcolor}

\begin{document}
\title{Computational Microstructure Analysis of Sintered Ceramics}
%
%
\author{
Nikhil~Dhanankam\inst{1}\orcidID{0009-0003-4331-6152} \and 
Deeksha~Kodangal\inst{3}\orcidID{0000-0001-5766-8026} \and
Rajendra~K.~Bordia\inst{3}\orcidID{0000-0001-9256-0301} \and
Ulf~D.~Schiller\inst{1,2}\orcidID{0000-0001-8941-1284}
}
\authorrunning{N. Dhanankam et al.}
%
\institute{
Department of Computer and Information Sciences, University of Delaware, Newark, DE 19716, USA \and
Department of Materials Science and Engineering, University of Delaware, Newark, DE 19716, USA \and
Department of Materials Science and Engineering, Clemson University, Clemson, SC 29634, USA\\
\email{\{nikhildh,uschill\}@udel.edu}
}

\maketitle

\begin{abstract}
Characterizing materials through manual extraction of physical properties from microstructure images is a laborious process. This work presents a workflow to extract porosity, solid fraction, grain size distribution, and pore size distribution from scanning electron microscopy (SEM) images of sintered ceramic samples using an automated pipeline. The primary challenge for extracting physical properties from SEM images is the presence of unimodal histograms in SEM images as a result of the overlapping intensity ranges for the grain and pore phases. We evaluated several different methods for noise reduction and local thresholding of SEM images. We find that topological filtering in combination with Sauvola thresholding enables segmentation and extraction of physical property data from SEM images. We validated the automated pipeline by comparing our results with the results of manual analyses performed for samples sintered at 1200$^o$C and 1400$^o$C and achieved an Intersection over Union (IoU) score of 95.14\% and 99.85\%, respectively. The workflow provides an efficient means to automatically extract microstructure properties from SEM images as a crucial step in generating materials datasets for machine learning.
\keywords{image analysis \and microstructure characterization \and additive manufacturing}
\end{abstract}

\section{Introduction}

Sintering-assisted additive manufacturing (SAAM) enables the fabrication of complex ceramic components for many different applications, including energy systems and hypersonic vehicles~\cite{Chen2023}. Traditional microstructure characterization methods require manual segmentation followed by extraction of physical characteristics of microstructure using software tools such as ImageJ~\cite{Schroeder2020}. As a result, the manual process achieves very low throughput rates, and the subjective nature of the segmentation process can create additional variability in the data produced. The computer vision community has developed numerous automated segmentation techniques. Sezgin and Sankur~\cite{Sezgin2004} reviewed over 40 different thresholding techniques and classified these techniques based on histogram shape, clustering, entropy object attributes, spatial correlation, and local adaptive methods. They concluded that there will not be one method that is optimal for all image types, and therefore, each specific image domain will require a different method that will produce optimal results for that specific domain. 
The time bottleneck has broader implications for the field of data-driven materials development and research, as machine learning models for materials science will use annotated datasets to identify and learn about processing-structure-property relationships~\cite{Schmidt2019,Agrawal2019}. However, because the experimental nature of materials research creates uncertainties and barriers to producing new data for use in machine learning model building, there will continue to be a fundamental limitation to the availability of suitable training data.
Global thresholding methods like Otsu's Method~\cite{Otsu1979} work well on images of bimodal histograms, which have a clear separation between the two phases. The challenge of conducting manual analysis of SEM images is caused by the type of microstructures found in sintered ceramics, which are characterized by unimodal distribution of grain and pore intensities within SEM images. This means that grain and pore intensities overlap, so they do not have distinct peaks or local minima or maxima on the histograms. Global thresholding methods are not appropriate for unimodal histograms when the intensity distributions of the image overlap significantly~\cite{Patel2022}. A new method for thresholding was proposed by Kapur et al.~\cite{Kapur1985} using an entropy-based metric to maximize the total entropy of the segmented regions. Niblack~\cite{Niblack1986} developed the first local adaptive thresholding method, based on the local mean and standard deviation of pixel values. Sauvola and Pietikäinen~\cite{Sauvola2000} improved on Niblack's approach by introducing a sensitivity parameter that enhances thresholding results for images with variable contrast.
Automated segmentation techniques have also been applied in materials science. Choi and Choi~\cite{Choi2001} applied watershed segmentation for the automated detection of grain boundaries in metallic microstructures. Morphological decomposition techniques have been proposed by Decencière and Jeulin~\cite{Decenciere2001} for studying the geometry of complex random structures in materials. Gostick et al.~\cite{Gostick2017} introduced an image-processing toolkit for porous media characterization. Most of these methods rely on the assumption of clearly defined grain boundaries between particles or lack of significant noise, which is not always valid for many ceramic systems. Meyer and Beucher~\cite{Meyer1990} developed the marker-controlled watershed algorithm, in which the seed markers can be positioned to provide more accurate segmentation. Adams and Bischof~\cite{Adams1994} developed algorithms for a seeded region growing technique, aggregating pixels based on their similarity.

\section{Materials and Methods}

\subsection{Image Dataset and Sample Preparation}

\begin{figure}
\centering
\includegraphics[width=0.46\textwidth]{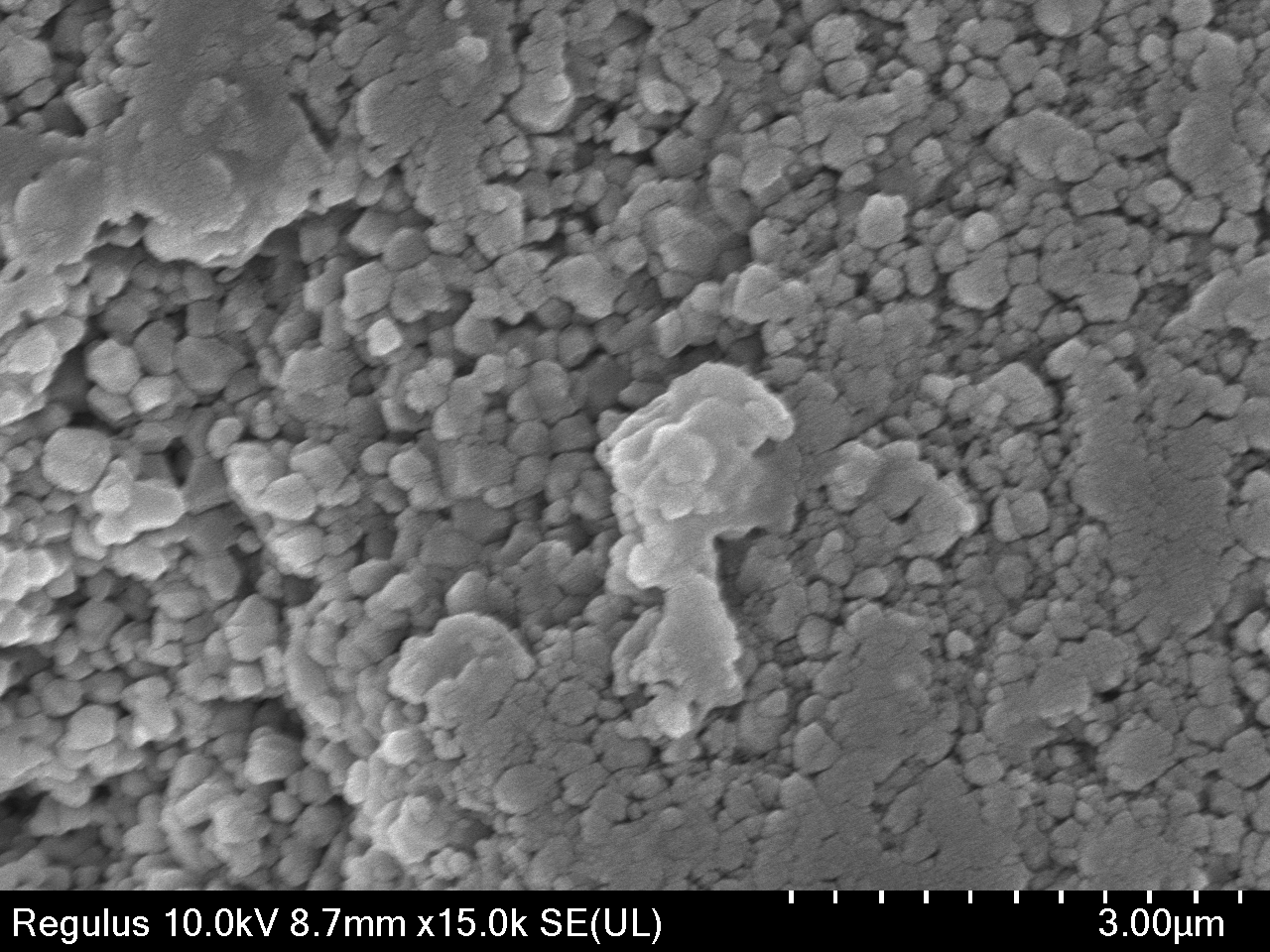}
\includegraphics[width=0.46\textwidth]{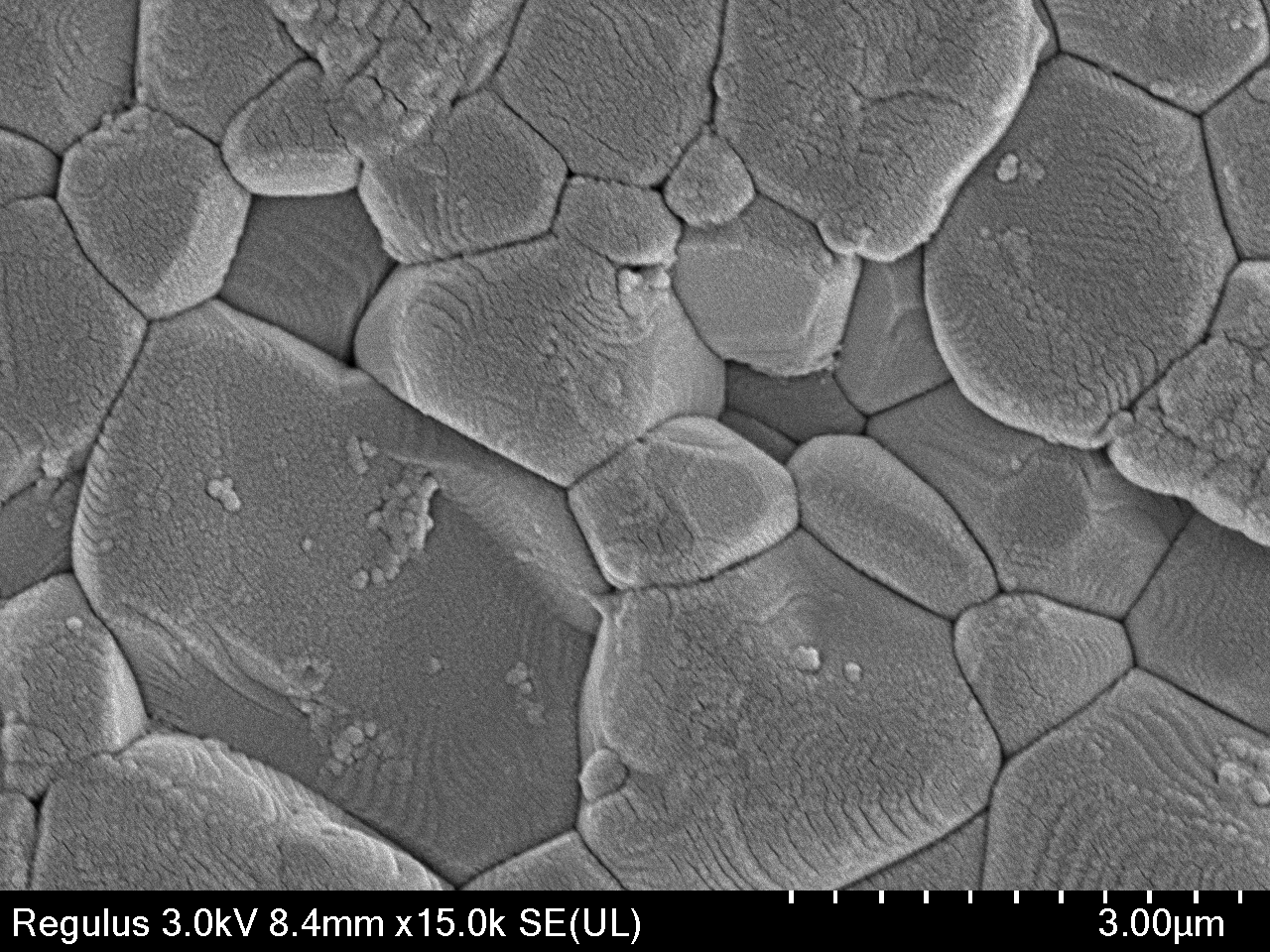}
\caption{Examples of SEM Images of slip-cast alumina samples sintered at (left) 1200$^{\circ}$C and (right) 1400$^{\circ}$C.\label{fig:sem_images}}
\end{figure}

An experimental dataset was collected from scanning electron microscopy (SEM) images taken from slip-cast samples made of alumina (Al$_2$O$_3$) that had been sintered at two different temperatures, one at 1200°C (169 images) and 1400°C (206 images). Green bodies were made using slip casting and were dried under a 90\% relative humidity, debound using a temperature of 600°C, and then sintered using a heating rate of 10°C per minute to their target temperatures, held for 2 hours, and allowed to cool afterwards.
The images were taken from a series of cylindrical cross sections at regular intervals, 0.1 mm apart in the horizontal direction and 0.5 mm in the vertical direction. The SEM used to take the images was a Regulus Scanning Electron Microscope and was able to take images with an accelerating voltage of 3.0-10.0 kV, an 8.4-8.5 mm working distance, and a magnification of 15,000 (with a scale bar of 3.00 μm). The imaging modality at this point was secondary electron (SE) and was set for ultra-low (UL) landing energy to show greater detail on the surface.
The sintering temperatures correspond to different stages of microstructure evolution. The samples heated to 1200°C exhibited a fine-grained porous structure where the individual grains are clearly visible, while at 1400°C, nearly complete densification occurred along with major grain growth and well-defined boundaries. Representative examples of the SEM images of the microstructures are shown in Fig.~\ref{fig:sem_images}. 

\subsection{Image Analysis Pipeline}

The Python programming language and standard libraries that support scientific calculations are the basis for developing the automation pipeline for the project. These include image processing library, OpenCV~\cite{Bradski2000}, the scientific computation library SciPy, and the image processing library scikit-image~\cite{VanDerWalt2014}. The pipeline includes the following five phases: 1) Noise Filtering, 2) Image Binarization, 3) Calculation of Solid Fraction and Porosity, 4) Grain Size and Distribution 5) Pore Size and Distribution. For comparing the results of the workflow, we use PoreSpy~\cite{Gostick2019}, an open source Python-based toolkit, which performs quantitative analysis of porous materials from volumetric images. Functionality in PoreSpy is divided into six categories: generators for creating synthetic images, filters for applying morphological operations, metrics for quantitative analysis, networks for creating pore networks, tools as helper functions, and io for converting formats.

\subsection{Noise Filtering}

\begin{figure}
\centering
\includegraphics[width=0.32\textwidth]{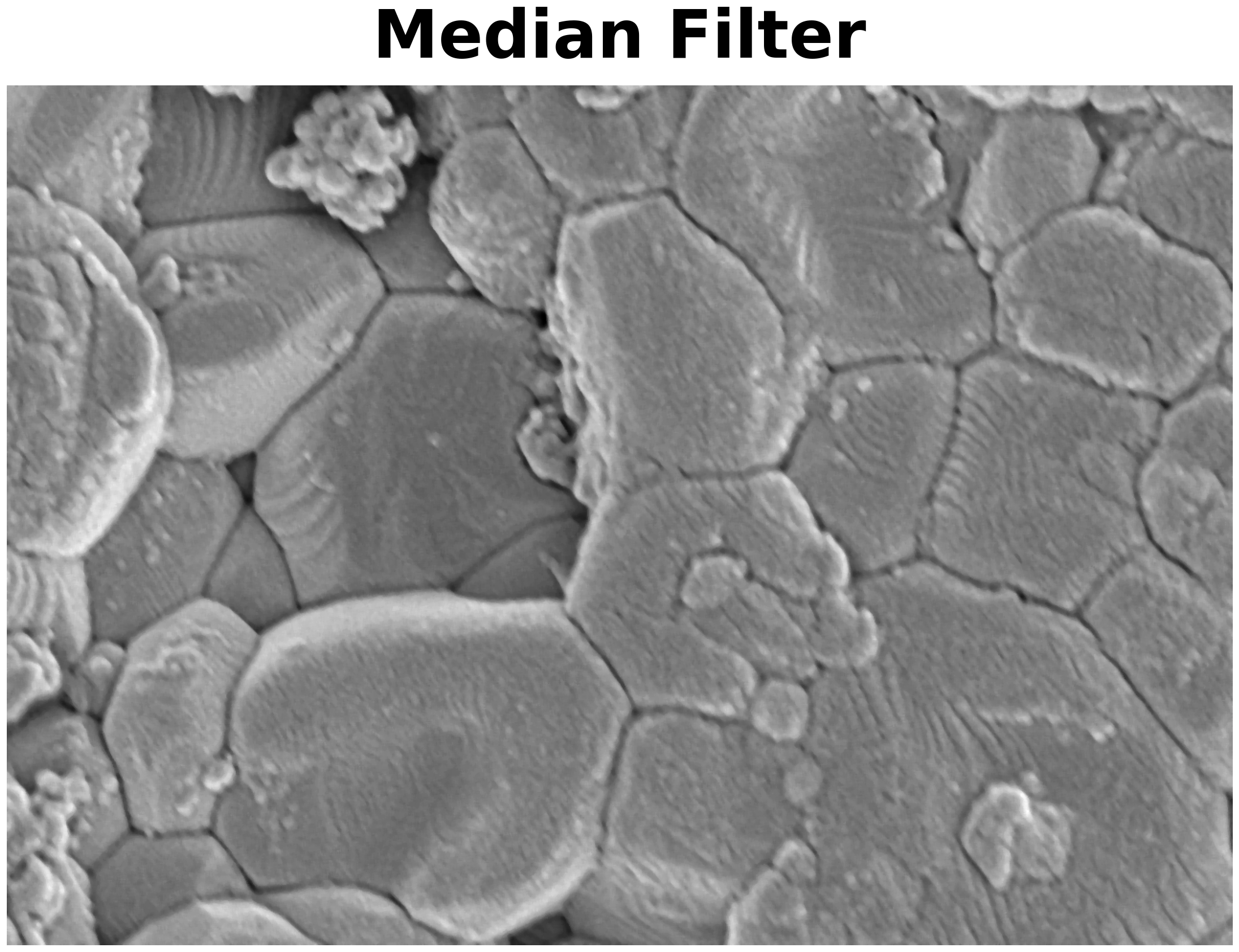}
\includegraphics[width=0.32\textwidth]{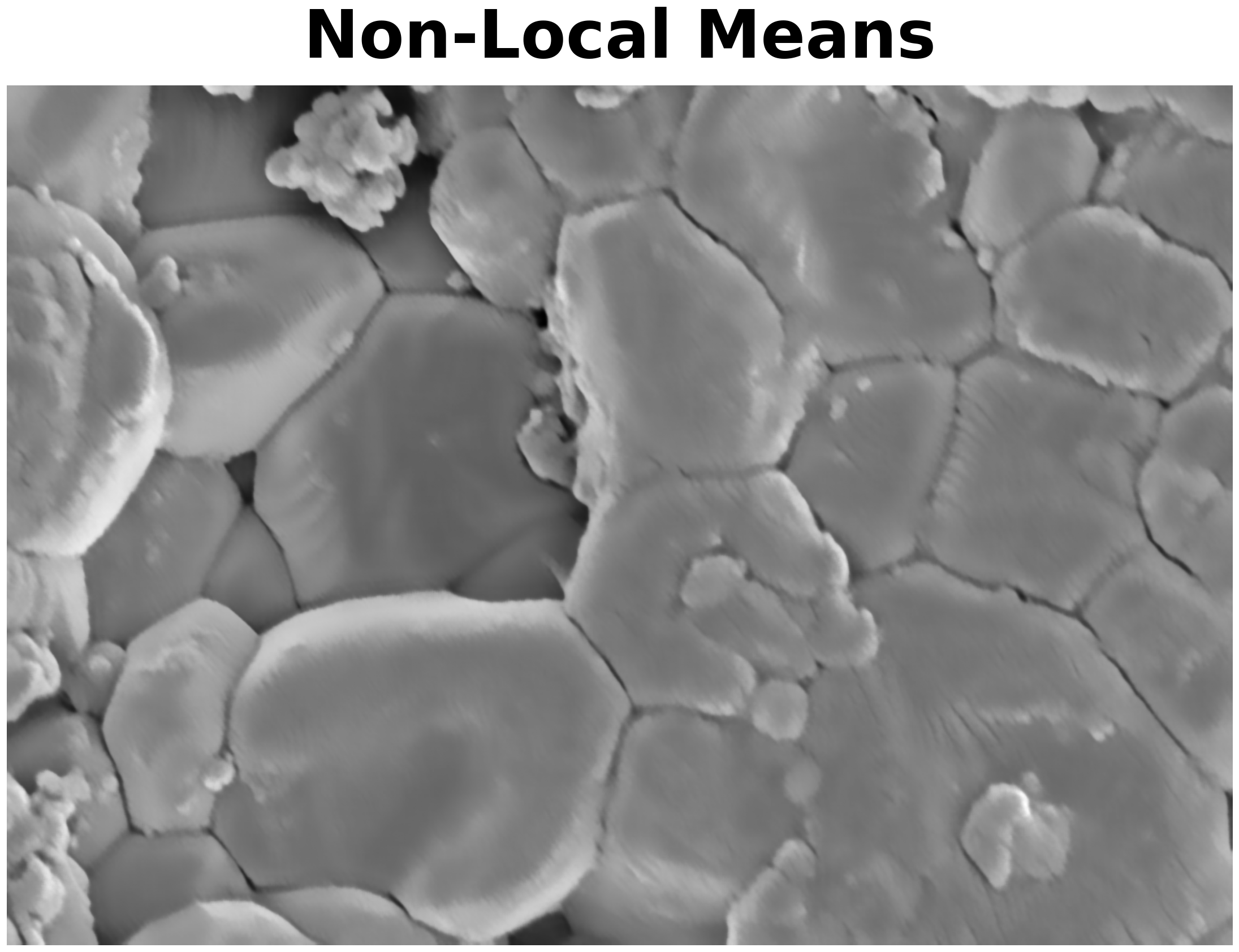}
\includegraphics[width=0.32\textwidth]{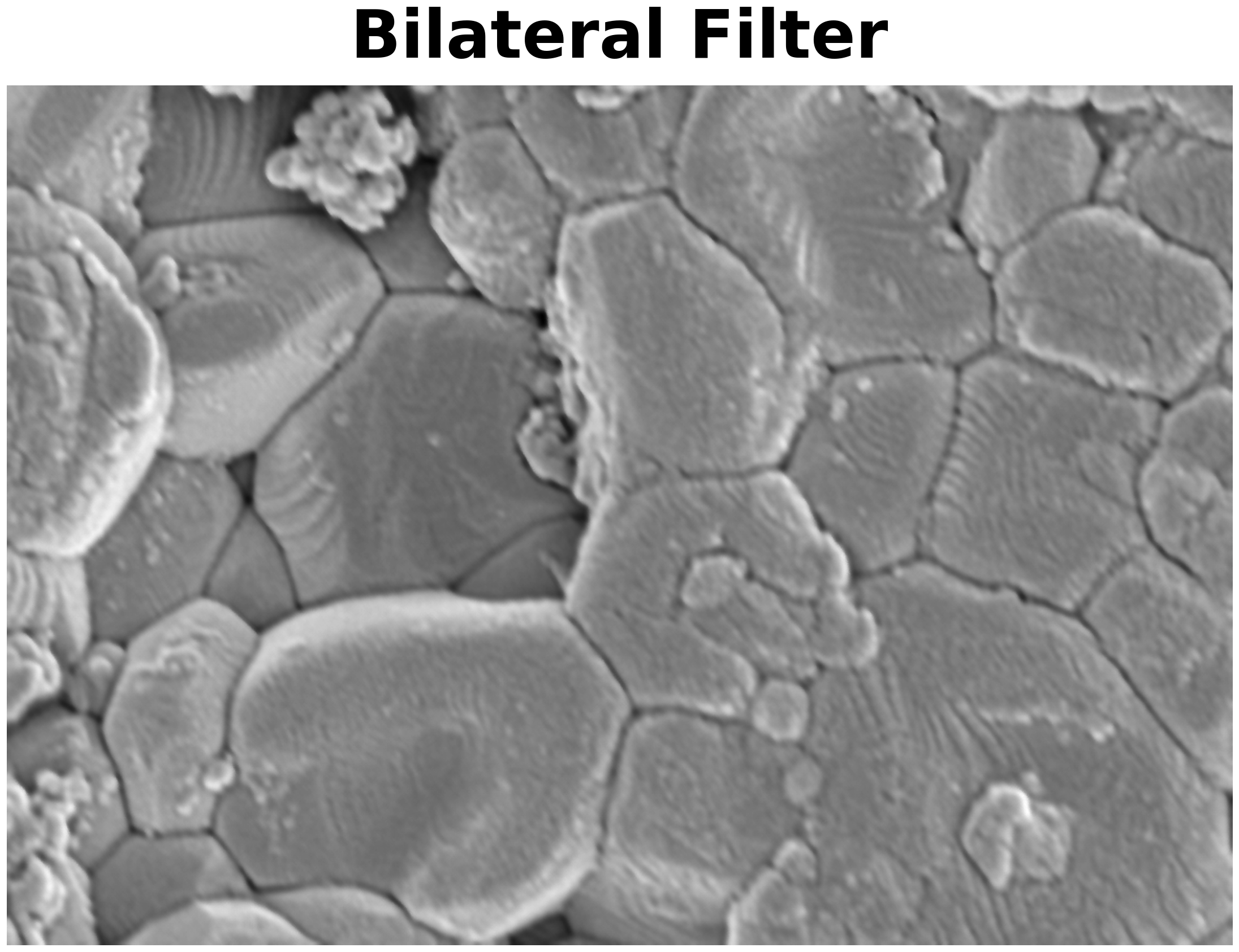}
\includegraphics[width=0.32\textwidth]{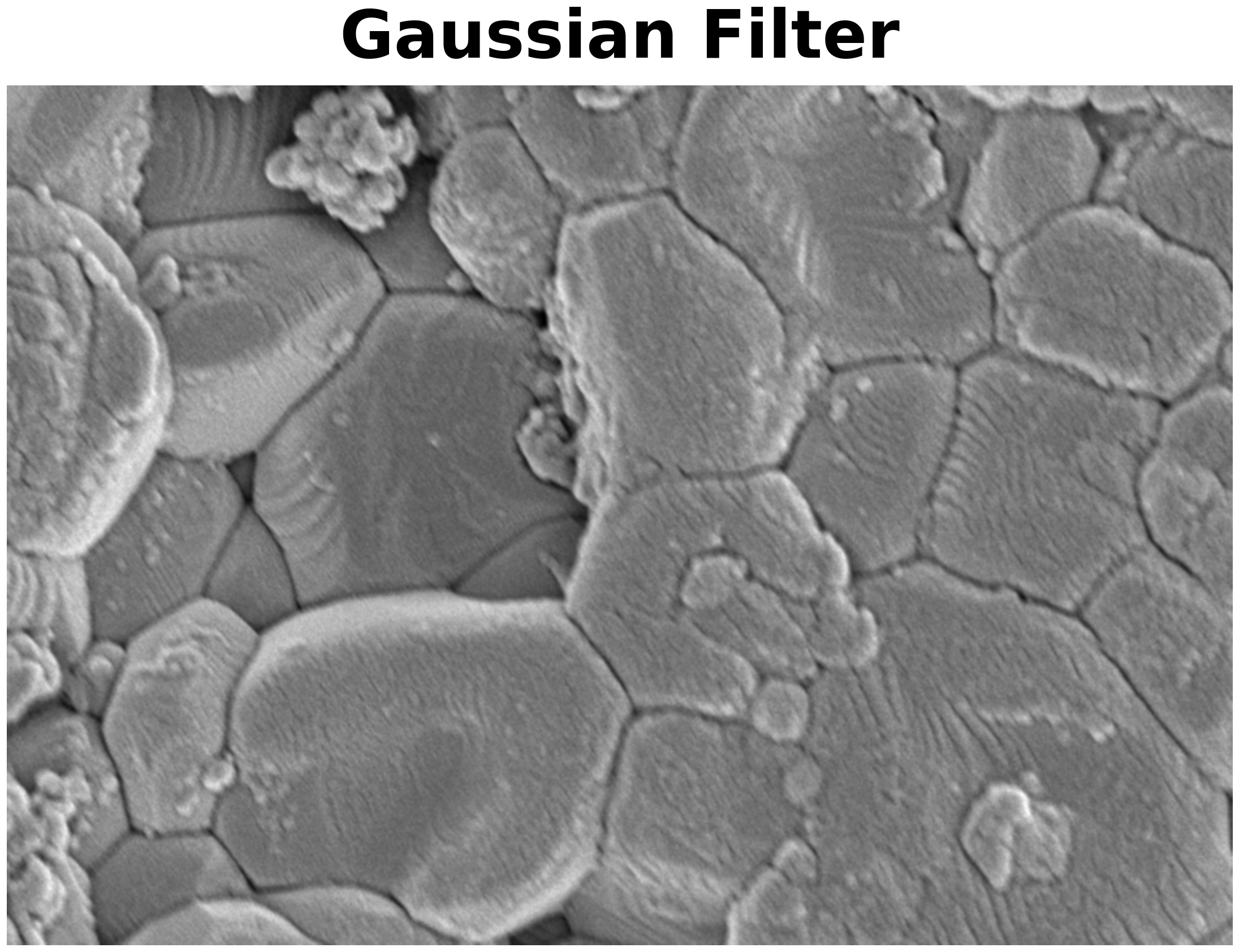}
\includegraphics[width=0.32\textwidth]{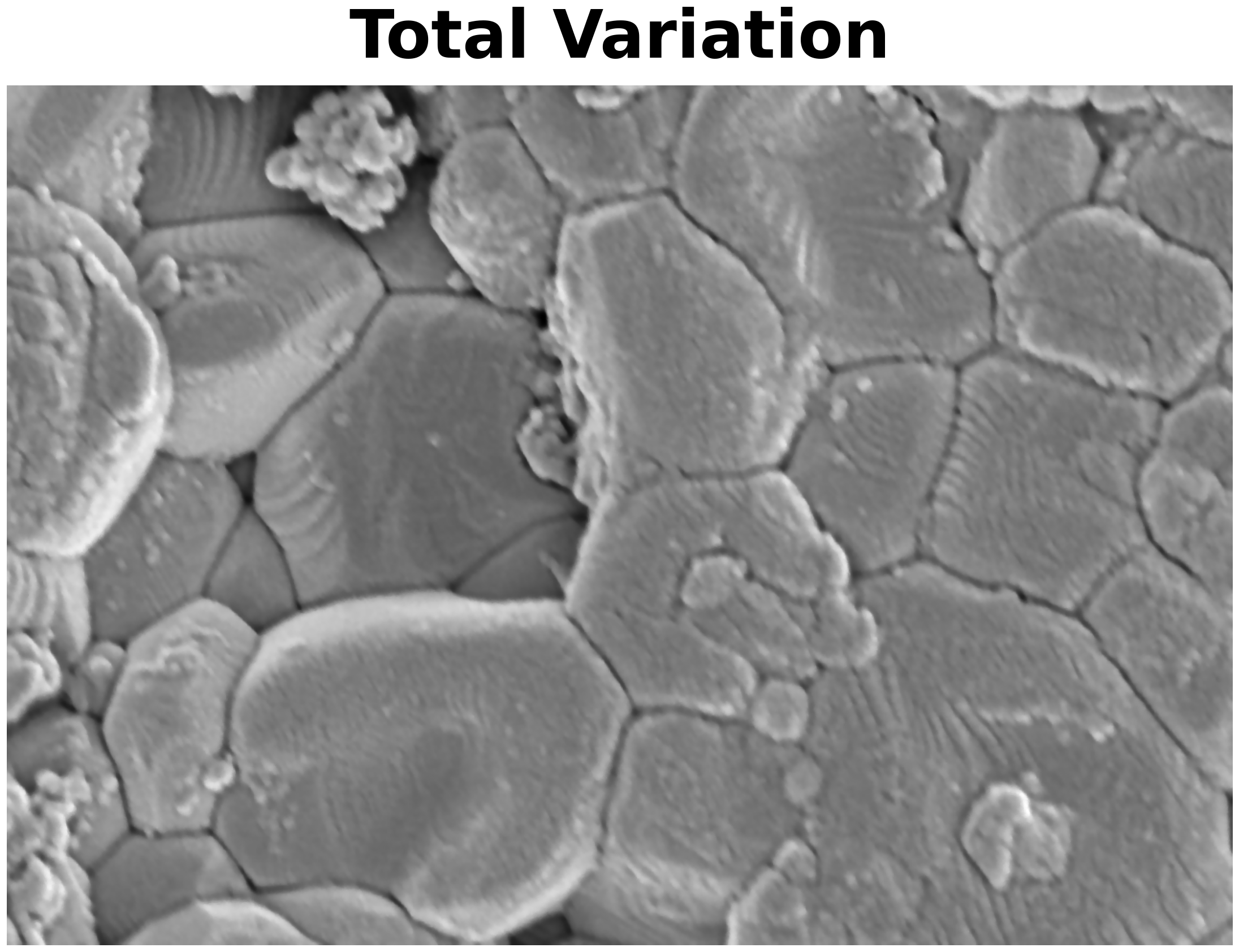}
\includegraphics[width=0.32\textwidth]{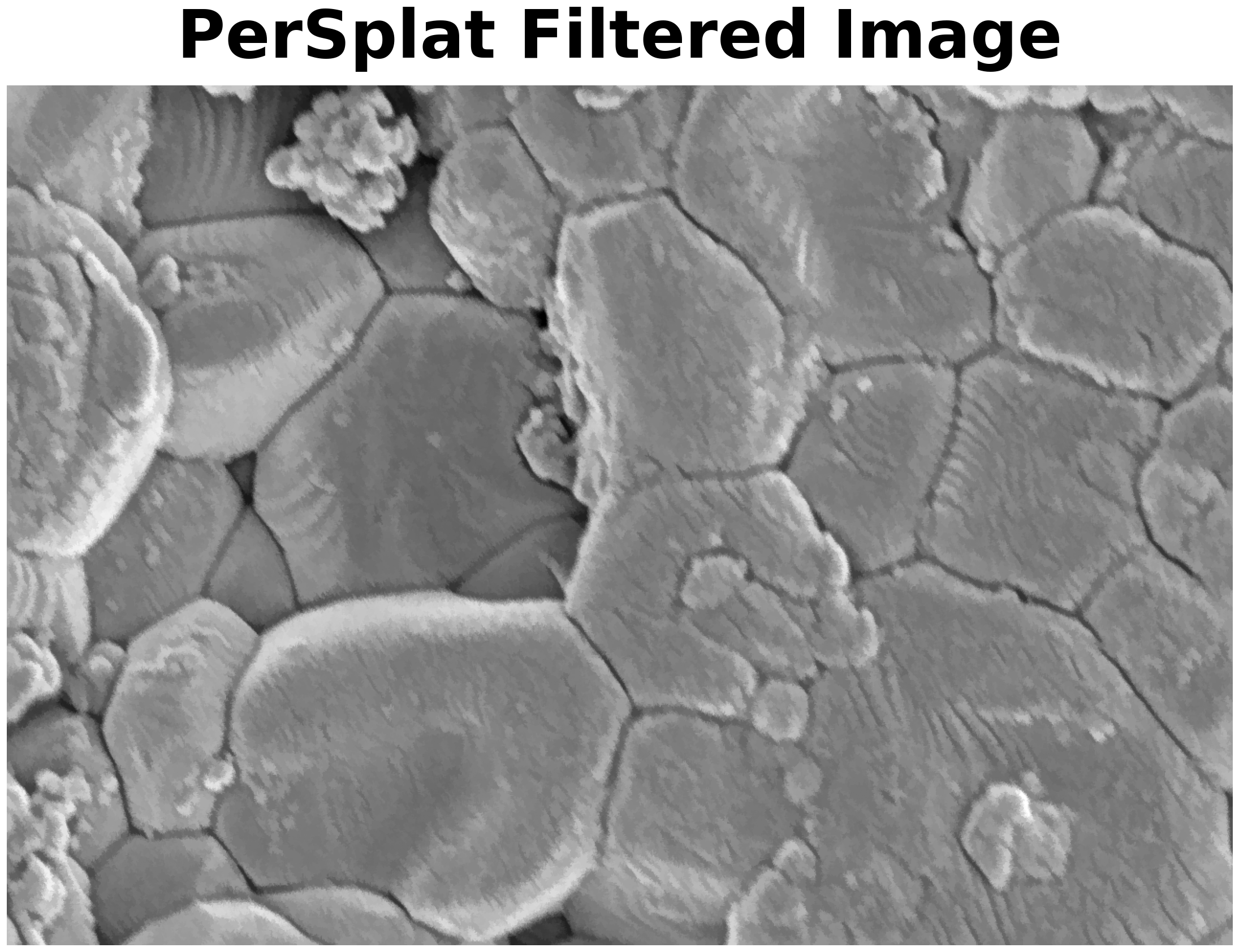}
\caption{Comparison of noise filtering algorithms.\label{fig:filtering}}
\end{figure}

SEM images suffer from shot noise, thermal noise, and charging effects on insulating surfaces~\cite{Goldstein2018}. These sources of noise degrade the image quality, which can be mitigated by noise filtering algorithms. We experiment with six filtering techniques across the four primary denoising paradigms~\cite{Gonzalez2018} to identify a filter to preserve microstructure. Spatial domain denoising reduces the noise by operating on the pixel intensity values of the original image. In Gaussian denoising, a linear filter is applied to smoothen the image. Median denoising~\cite{Huang1979} is a nonlinear edge-preserving filter, which eliminates salt-and-pepper noise. Bilateral denoising~\cite{Tomasi1998} is a modern edge-preserving filter that uses spatial filtering and range filtering to eliminate noise while preserving edges. In frequency domain denoising methods, the image is transformed into a frequency representation. The non-local means~\cite{Buades2005} denoising method takes advantage of similar textures that occur over multiple locations throughout the entire image and is particularly effective for repeated texture patterns (grain structures). Optimization-based denoising methods include total variation~\cite{Rudin1992}, i.e., minimizing total variation while also preserving edges. Topological denoising methods focus on preserving geometrical and structural features. PerSplat~\cite{Patel2022} uses persistent homology, which preserves topologically significant features and splats the noise. The evaluation of the six filtering methods included comparisons of peak signal-to-noise ratio (PSNR) and preservation of physical properties (porosity) with respect to the values obtained from manual segmentation in ImageJ for each filtering technique. Examples of filtered images are shown in Fig.~\ref{fig:filtering}.

\subsection{Image Binarization}

Sintered microstructure images exhibit unimodal histograms, i.e., grain and pore phases share overlapping intensity ranges without distinct peaks~\cite{Patel2022}. To overcome this, we use Sauvola thresholding based on the following three factors. First, from the theoretical concept, Sauvola uses the local mean and standard deviation to derive the statistics required for microstructures with an overlapping intensity distribution. Methods like Bernsen~\cite{Bernsen1986}, which uses the maximum and minimum range, are ineffective when local contrast is low, as in densified ceramic materials. Second, prior applications in materials science illustrate Sauvola has already been used successfully on low-contrast biomedical images~\cite{Ronneberger2015} and metallography~\cite{Chowdhury2016}.
Sauvola thresholding computes local threshold as
\begin{equation}
T(x,y) = \mu(x,y)  \left[ 1  + k \left( \frac{\sigma(x,y)}{R} - 1 \right) \right] ,
\label{eq:sauvola}
\end{equation}
where $x$ and $y$ are the pixel coordinates, $\mu(x,y)$ and $\sigma(x,y)$ are the local mean and standard deviation computed over a square window size $w\times w$ centered at pixel $(x,y)$, $k$ is a sensitivity parameter that governs how much the local standard deviation pulls the threshold away from the local mean, and $R$ is the dynamic range of the standard deviation. While $R$ has minimal influence on the binarization quality, both $k$ and $w$ substantially influence segmentation quality and requires image-specific tuning~\cite{hadjadj2016isauvola}. Local hyperparameter optimization was performed, tuning the $k$ parameter and window size $w$ for each temperature class to compare against manually thresholded images.

\subsection{Solid Fraction, Porosity, Grain Size and Pore Size Calculation}

\begin{figure}
\centering
\includegraphics[width=0.38\textwidth]{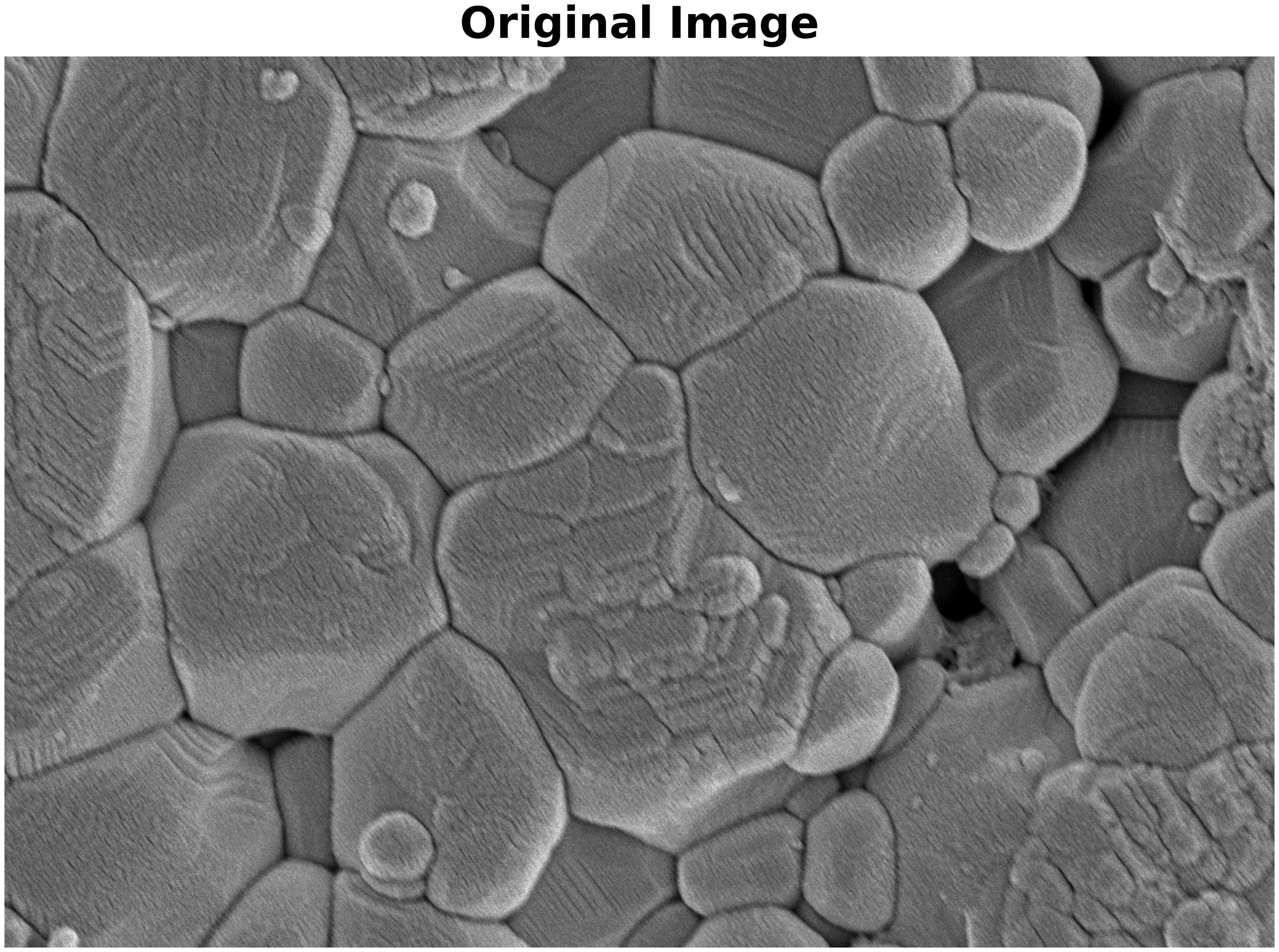}
\includegraphics[width=0.38\textwidth]{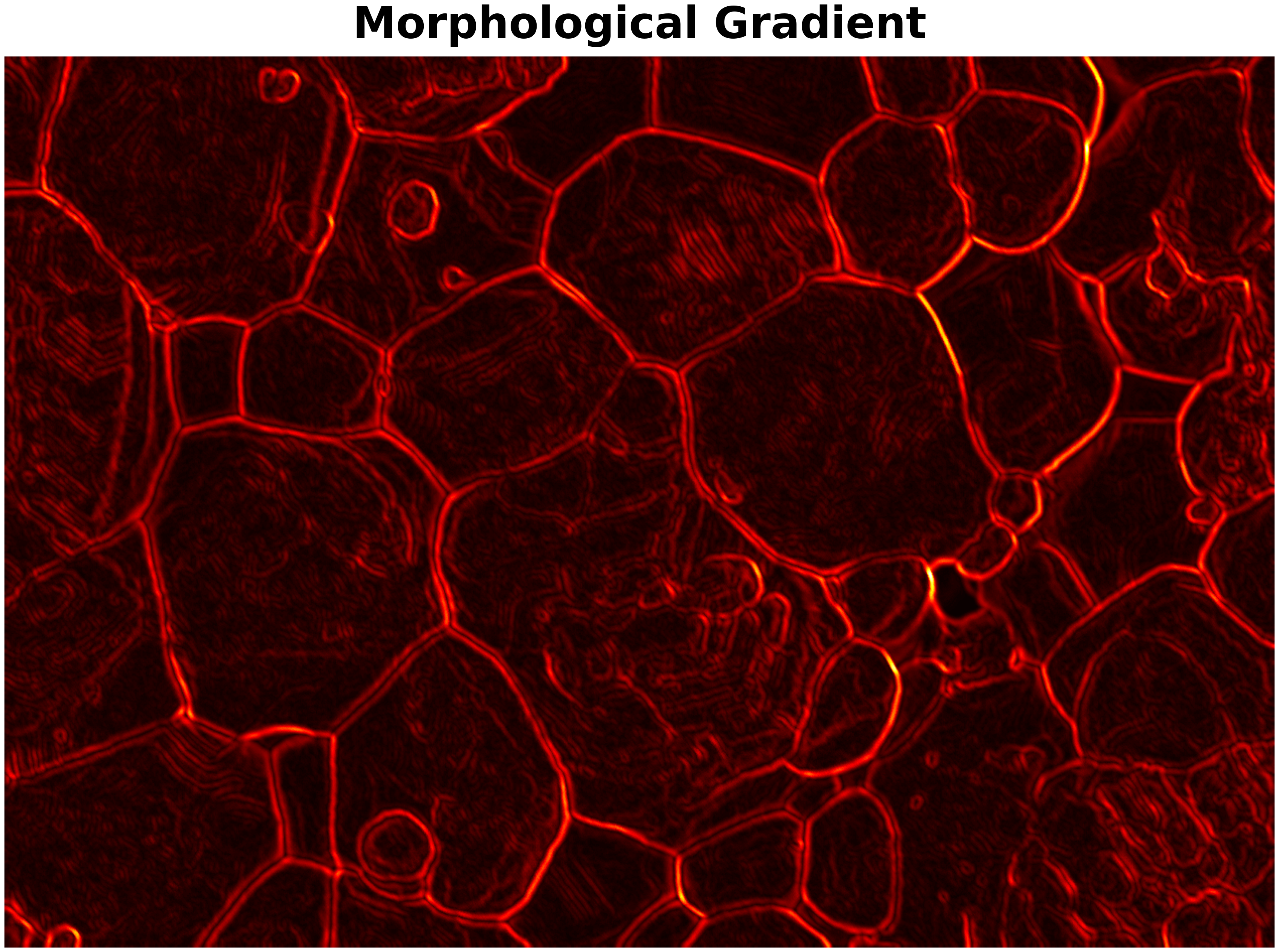}
\includegraphics[width=0.38\textwidth]{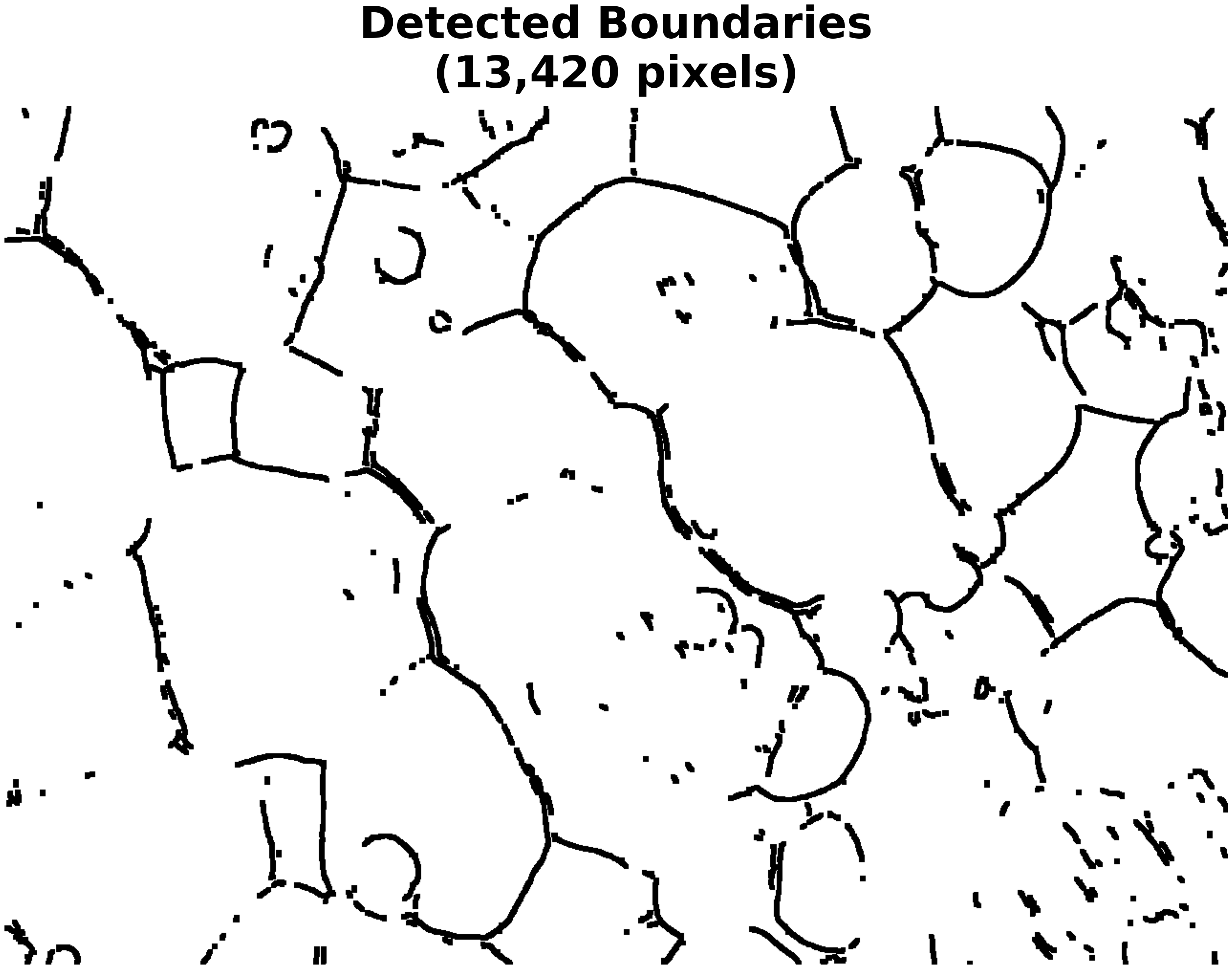}
\includegraphics[width=0.38\textwidth]{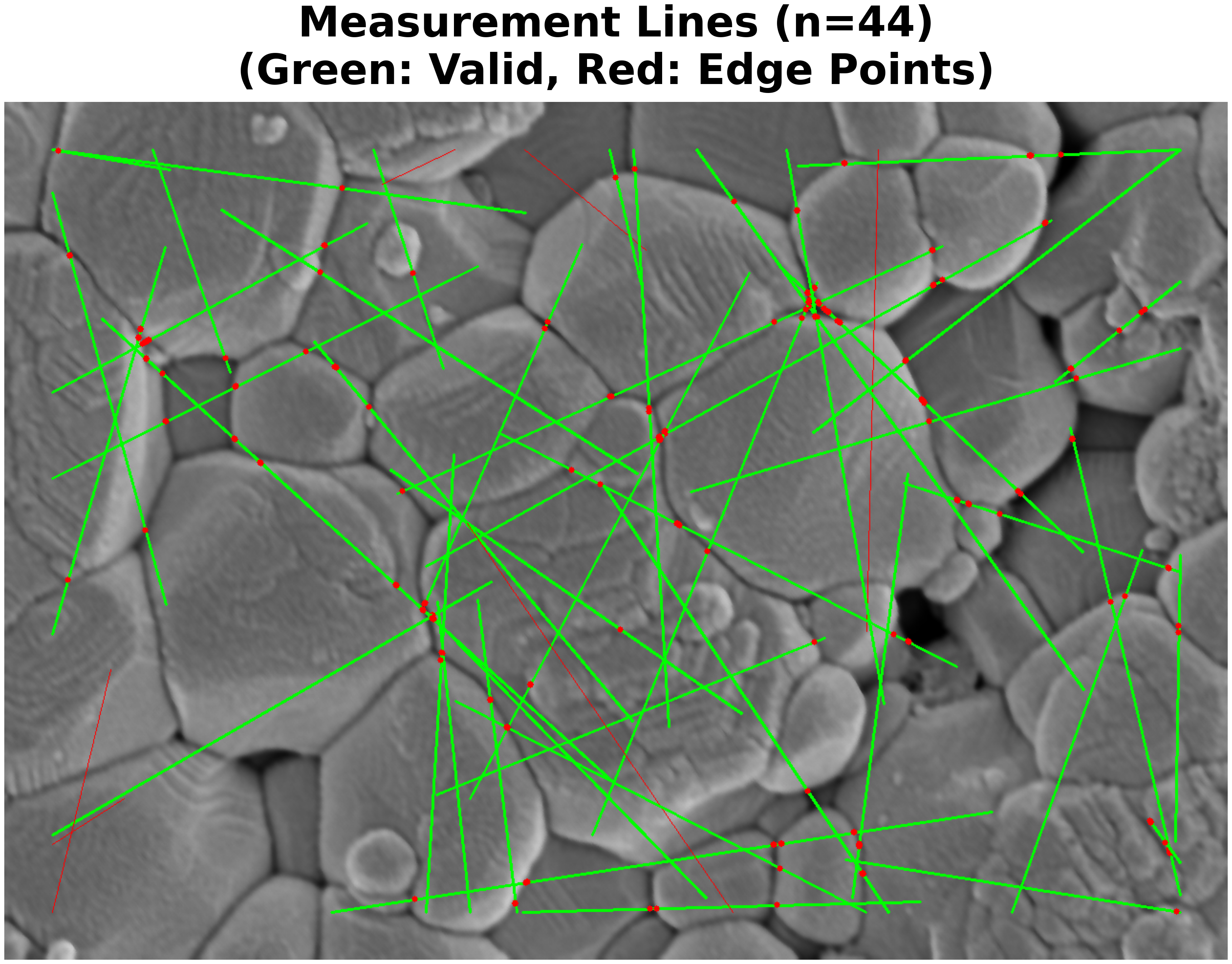}
\caption{Grain size analysis workflow: (top-left) original image, (top-right) morphological gradient, (bottom-left) detected boundaries, and (bottom-right) measurement lines with edge points.\label{fig:grain_size}}
\end{figure}

Following binarization, the solid fraction $\varphi$ is calculated as
\begin{equation}
\varphi = \frac{N_{\mathrm{white}}}{N_{\mathrm{total}}} ,
\label{eq:solid_fraction}
\end{equation}
where $N_\text{white}$ is the number of white pixels and $N_\text{total}$ is the total number of pixels. Porosity $P$ is the complement
\begin{equation}
P = 1 - \varphi .
\label{eq:porosity}
\end{equation}
Grain size analysis follows the four-step workflow illustrated in Fig.~\ref{fig:grain_size}: (1) Bilateral filtering for preprocessing, (2) morphological gradient for boundary detection, (3) skeletonization for boundary thinning, and (4) random measurement lines for mean intercept length calculation~\cite{ASTM2013}, providing an estimate of average grain size.
Bilateral filtering was used to remove noise while preserving edges, the morphological gradient was calculated with a 5×5 kernel to highlight grain boundaries, and adaptive thresholding was used to develop a binary map of grain boundaries. The thresholding values used in the binary map were 20 for the 1200°C test and 35 for the 1400°C. The random line intercept method was applied by generating $n=50$ random measurement lines across each image~\cite{Abrams1971}, detecting the number of intersections along each line with grain boundaries, and filtering out noise using a minimum gap filter of 3 pixels for 1200°C and 12 pixels for the 1400°C images. The number of randomly measured lines was selected based on recommendations for measuring lognormal grain size distributions of small dataset sizes (n<100)~\cite{Lopez-Sanchez2020}. Mean grain size and standard deviation were computed across the whole dataset for each temperature class.

\begin{figure}
\centering
\includegraphics[width=0.38\textwidth]{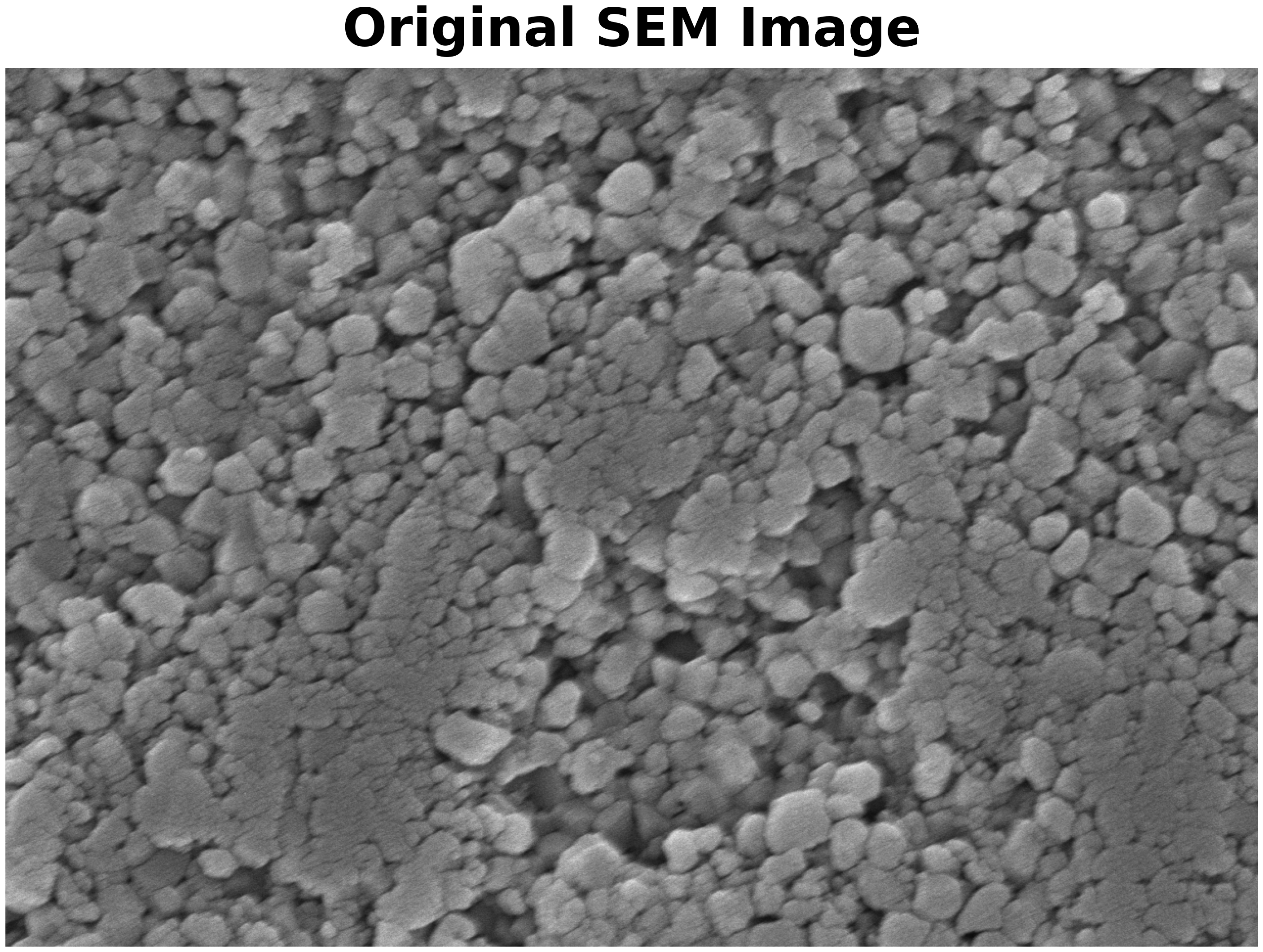}
\includegraphics[width=0.38\textwidth]{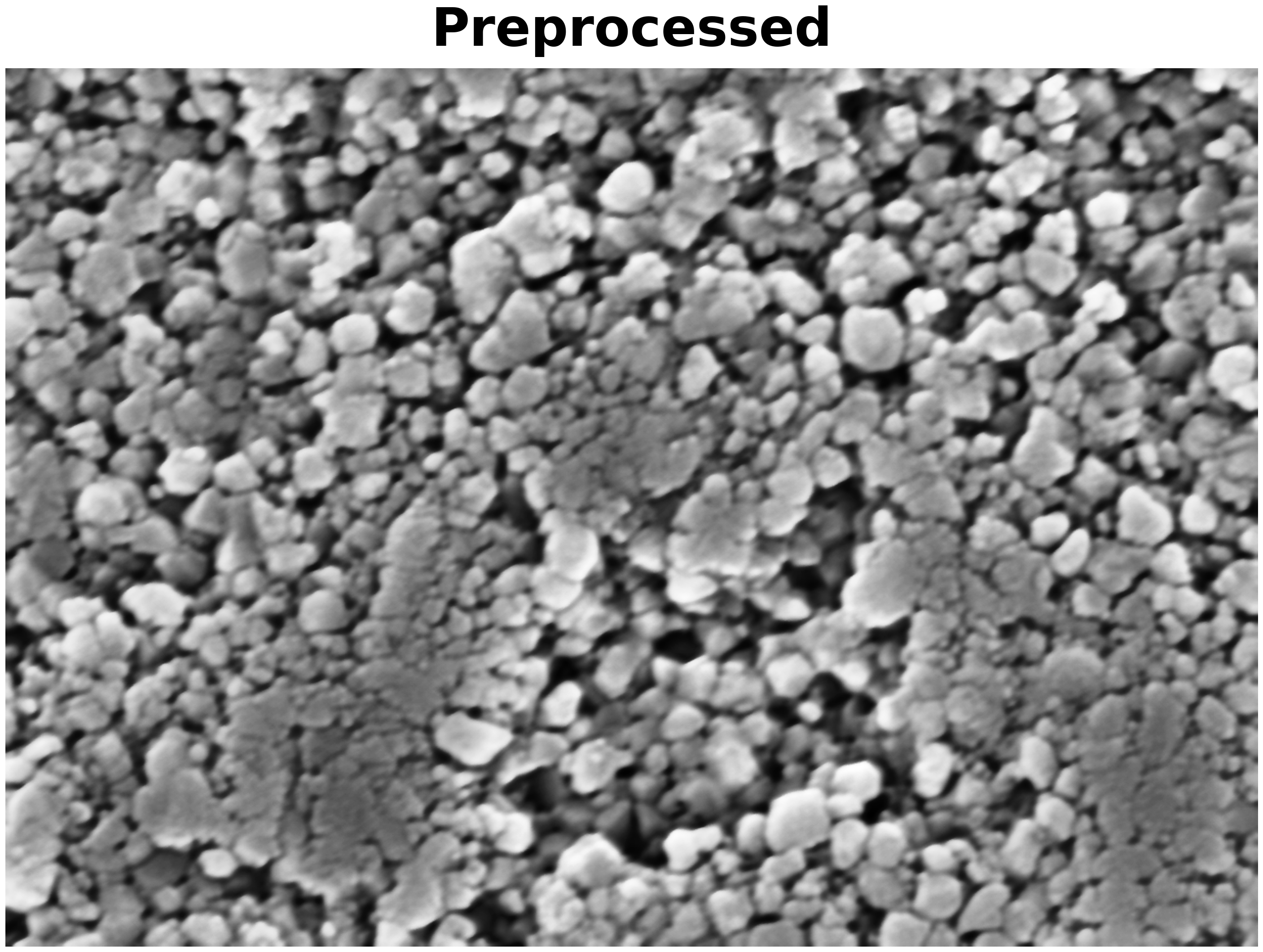}
\includegraphics[width=0.38\textwidth]{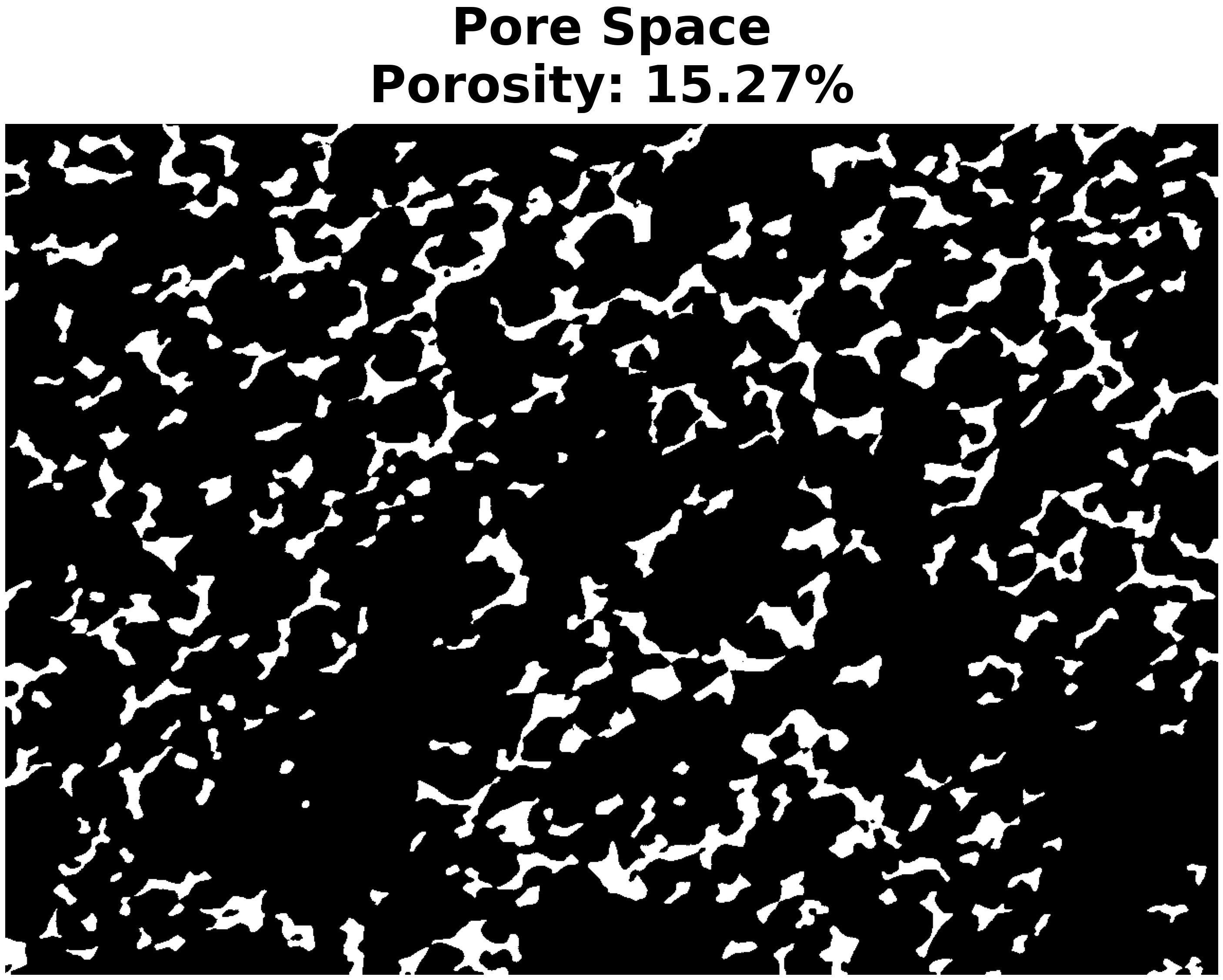}
\includegraphics[width=0.38\textwidth]{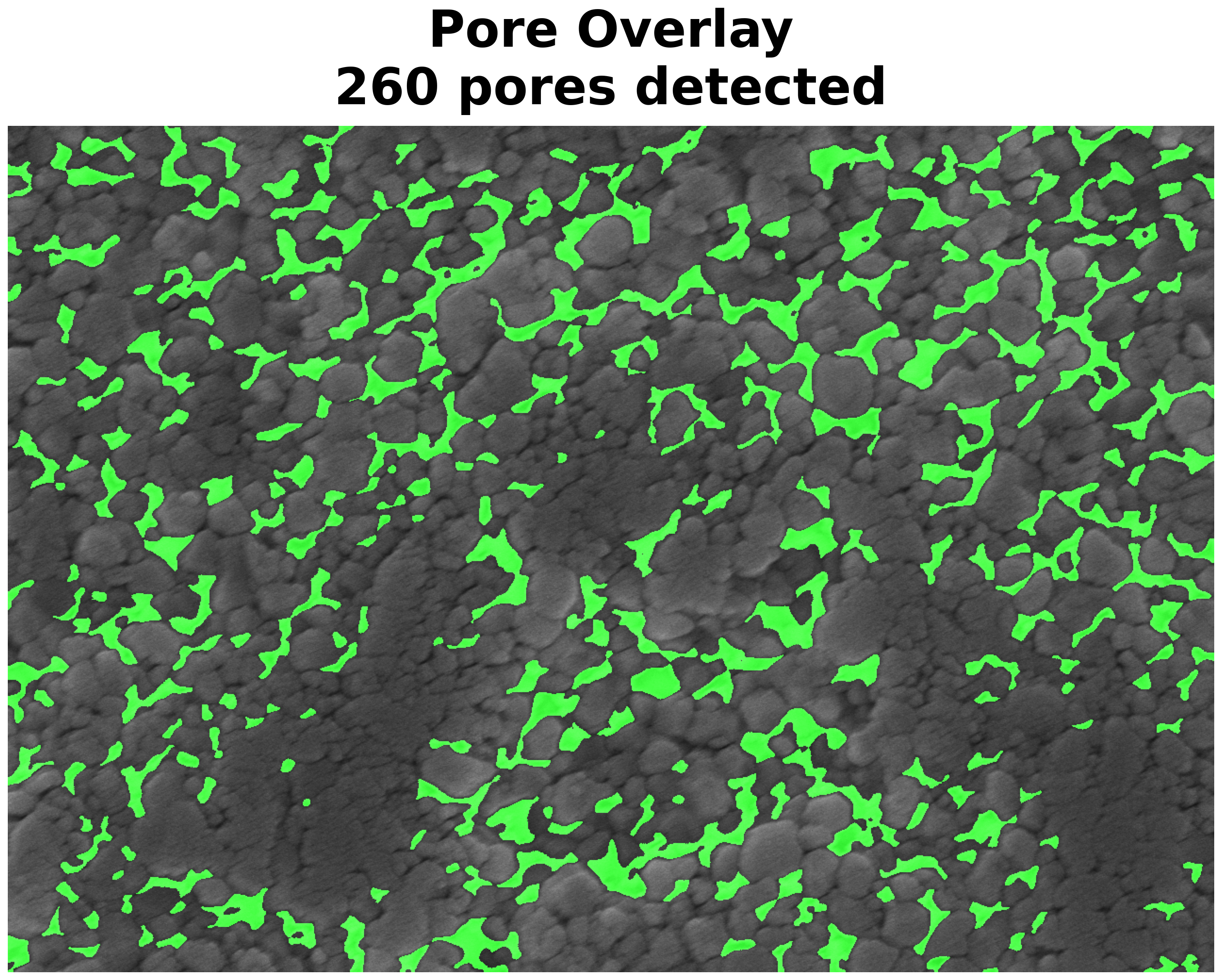}
\caption{Processing steps for computation of pore size distribution: (top row, left to right) Original SEM image, preprocessed image with bilateral filtering and CLAHE, final pore mask; (bottom row) pore overlay showing detected pores, and individual pores color-coded for analysis.\label{fig:pore_workflow}}
\end{figure}

Pore detection utilizes morphological reconstruction using dual thresholding~\cite{Girard2011}. This includes two complementary masks, the Sauvola threshold to detect all pore boundaries, and a percentile threshold that detects definite pore regions. The workflow begins with bilateral filtering followed by contrast-limited adaptive histogram equalization (CLAHE)~\cite{Zuiderveld1994} with a clip limit of 2.0 and a tile size of 8×8 pixels to enhance local contrast.
The Sauvola boundary mask utilizes adaptive thresholding to identify all potential pore pixels. Combined with a percentile seed mask, which identifies definite pore regions using pixels below the 5th percentile of image intensity. True pore regions were reconstructed through iterative geodesic dilation~\cite{Vincent1993,Soille2003}.
\begin{equation}
R^{(n)}_I(J) = (R^{(n-1)}_I(J) \oplus B) \wedge I
\end{equation}
where $J$ is the seed mask, $I$ is the boundary mask, $B$ is the structuring element, $\oplus$ denotes dilation, and $\wedge$ denotes the point-wise minimum. Iteration continues until idempotency is reached~\cite{Vincent1993}.
\begin{equation}
(I \oplus B)(x) = \max_{b \in B} \, I(x - b)
\end{equation}
The reconstruction expands the conservative seeds to fill accurate pore shapes while eliminating false positives. For each pore, we compute the area, perimeter, equivalent diameter. An example of the pore detection workflow is shown in Fig.~\ref{fig:pore_workflow}.

\subsection{Evaluation Metrics}

\subsubsection{Intersection over Union (IOU)}

The intersection-over-union (IOU) metric, also known as the Jaccard similarity coefficient~\cite{Jaccard1912}, measures the overlap between predicted segmentation $P$ and the ground truth segmentation $G$.

\begin{equation}
\mathrm{IoU} = \frac{|P \cap G|}{|P \cup G|}
\label{eq:iou}
\end{equation}
The use of the IOU metric as a segmentation benchmark is a long-established practice for computer vision tasks~\cite{Rahman2016,Csurka2013}. While pixel accuracy can be misleading when datasets are imbalanced, this is not true for IOU, which equally penalizes false positive and false negative results~\cite{Csurka2013}. This distinction is important for our application since pores only occupy a small fraction (2\%-12\%) of the image area. IOU has been used as a metric for segmentation accuracy in the field of material research. For instance, DeCost et al.~\cite{DeCost2019} utilized IOU to validate the segmentation of steel microstructures through deep learning techniques. Azimi et al.~\cite{Azimi2018} reported IOU for automated steel phase segmentation, and Chowdhury et al.~\cite{Chowdhury2016} used the Jaccard index (IOU) for microstructure classification.

\subsubsection{Signal-to-Noise Ratio (SNR)}

The Signal-to-noise ratio (SNR) measures the quality of filtered images as
\begin{equation}
\mathrm{SNR} = 10 \log_{10}\!\left(\frac{P_{\mathrm{signal}}}{P_{\mathrm{noise}}}\right) ,
\label{eq:snr}
\end{equation}
where $P_{\mathrm{signal}}$ is the power of the filtered image and $P_{\mathrm{noise}}$ is the power of the residual between the original and the filtered image, i.e.,
\begin{equation}
P_{\mathrm{signal}} = \frac{1}{N} \sum_{i=1}^{N} 
\left[ I_{\mathrm{filtered}}(i) \right]^2 ,
\label{eq:psignal}
\end{equation}
\begin{equation}
P_{\mathrm{noise}} = \frac{1}{N} \sum_{i=1}^{N} 
\left[ I_{\mathrm{original}}(i) - I_{\mathrm{filtered}}(i) \right]^2 .
\label{eq:pnoise}
\end{equation}
Here, $I_{\mathrm{original}}$ is the raw SEM image, $I_{\mathrm{filtered}}$ is the filtered image, and $N$ is the total number of pixels.In SEM images, signal refers to the actual differences in intensity produced by the microstructural features, while noise is the residual between the filtered and original images~\cite{Goldstein2018}. SNR has been a widely used metric for measuring the quality of images in signal processing~\cite{Gonzalez2018}. SNR has been used to evaluate filtering techniques for material characterization. DeCost et al.~\cite{DeCost2019} used SNR to compare preprocessing methods for metallography. Münch and Holzer~\cite{Munch2008} used SNR to evaluate noise reduction for tomographic reconstruction.

\section{Results}

\subsection{Noise Filtering Performance}

Traditional filtering techniques apply signal processing methods to analyze microstructures, whereas the PerSplat algorithm derives feature identification based on topological persistence on multiple levels of grayscale intensities.
Table~\ref{tab:noise-filtering} provides a summary of each filter on each temperature class. PerSplat performed well with an SNR increase of 25.0 dB for 1200°C and 22.3 dB for 1400°C images, yielding the porosity value closest to the manual segmentation.



\begin{table}
\centering
\caption{Comparison of signal-to-noise ratio (SNR) and computed porosity values for the filter algorithms applied to images of microstructures sintered at 1200$^{\circ}$C and 1400$^{\circ}$C.\label{tab:noise-filtering}}
\begin{tabular}{lcccc}
\hline
& \multicolumn{2}{c}{\textbf{1200$^{\circ}$C}} & \multicolumn{2}{c}{\textbf{1400$^{\circ}$C}} \\
\hline
Filter Method & SNR (dB) & Porosity & SNR (dB) & Porosity \\
\hline
Manual Segmentation & & 0.09 $\pm$ 0.02 & & 0.0027 $\pm$ 0.0019 \\
NL-Means & 22.3 & 0.05 $\pm$ 0.01 & 20.6 & 0.0004 $\pm$ 0.0005 \\
Gaussian & 17.7 & 0.04 $\pm$ 0.01 & 19.9 & 0.0003 $\pm$ 0.0004 \\
Total Variation & 17.0 & 0.03 $\pm$ 0.01 & 18.5 & 0.0002 $\pm$ 0.0004 \\
Median & 11.7 & 0.04 $\pm$ 0.01 & 14.7 & 0.0003 $\pm$ 0.0005 \\
Bilateral & 18.4 & 0.03 $\pm$ 0.01 & 19.0 & 0.0002 $\pm$ 0.0003 \\
PerSplat & 25.0 & 0.03 $\pm$ 0.01 & 22.3 & 0.0044 $\pm$ 0.0029 \\
\hline
\end{tabular}
\end{table}


\subsection{Binarization Validation}

Unlike materials with distinct phase contrast, sintered single-phase ceramics exhibit unimodal intensity distributions where grains and pores overlap in grayscale images. Sauvola thresholding works because it uses local image statistics rather than assuming global intensity bimodality. Results of the binarization and segmentation using Sauvola thresholding are reported in Table~\ref{tab:IOU-Comparison}. Sauvola thresholding achieved good agreement with manual analysis with a mean 92.18\% IOU for 1200°C images and 99.35\% IOU for 1400°C images.

\begin{table}
\centering
\caption{Comparison of segmentation using automated thresholding against manual segmentation using ImageJ. Values reported are Intersection over Union (IOU) scores.\label{tab:IOU-Comparison}}
\begin{tabular}{lccccc}
\hline
\multicolumn{6}{c}{\textbf{Segmentation Validation (IOU)}} \\
\hline
Temp. & Mean (\%) & Median (\%) & Std Dev (\%) & Min (\%) & Max (\%) \\
\hline
1200$^{\circ}$C & 92.18 & 92.82 & 2.24 & 86.07 & 95.14 \\
1400$^{\circ}$C & 99.35 & 99.45 & 0.35 & 98.64 & 99.85 \\
\hline
\end{tabular}
\end{table}

\subsection{Physical Property Extraction}

Results for the physical property extraction are shown in Table~\ref{tab:physical-properties}.
The 1200$^{\circ}$C class produced an average porosity of 12.4 ± 2.7\%. For 1400$^{\circ}$C, the porosity decreased to 2.44 ± 0.68\% , showing nearly complete densification.
The mean grain size increased by 3.7 times between 1200$^{\circ}$C and 1400$^{\circ}$C, which increased the atomic mobility and grain boundary migration as the temperature increased. We compared our grain size measurements with the NIST Grain Size Analysis Tool, an open-source software that allows users to automate grain boundary segmentation and measure grain sizes using ASTM E112 compliant intercept methods on SEM images~\cite{Moser2022}. The comparison with manual segmentation showed that our workflow is in closer agreement than GSAT.
The discrepancy is ascribed to the fact that NIST GSAT does not apply a minium gap filter between successive grain boundaries, and therefore overestimates the number of intercepts per line, resulting in a lower mean intercept length. In contrast, our workflow includes a minimum gap filter between consecutive boundary crossings, which reduces the number of spurious intercepts caused by noise artifacts, sub-grain contrast differences and resolved boundaries.
The morphological reconstruction used for pore identification accurately distinguishes true pores from grain boundaries. For the 1200$^{\circ}$C class, the  workflow identified an average of 292 $\pm$ 102 pores/image with a mean equivalent diameter of 0.148 $\pm$ 0.032 $\mu$m and a mean pore area of 0.023 $\pm$  0.008 $\mu$m$^2$, while for the 1400$^{\circ}$C class, 183 $\pm$ 61 pores/image were detected with a smaller average equivalent diameter of 0.074 $\pm$ 0.008 $\mu$m and an average pore area of 0.007 ± 0.002 $\mu$m$^2$. For comparison of our pipeline, we used PoreSpy~\cite{Gostick2019}. PoreSpy measured lower porosity values and smaller pore sizes than our workflow. The quantitative difference is due to PoreSpy's method of determining chord length distribution, which measures pore space statistically using random linear sampling rather than identifying discrete pore boundaries of individual pores. Our morphological reconstruction approach provides individual pore segmentation, which enables direct measurement of pore-specific properties.

\begin{table}
\centering
\caption{Physical property extraction results for microstructures sintered at 1200$^{\circ}$C and 1400${^\circ}$C.\label{tab:physical-properties}}
\begin{tabular}{lcccc}
\hline
\multicolumn{5}{c}{\textbf{Grain Size}} \\
\hline
& \multicolumn{2}{c}{1200$^\circ$C} & \multicolumn{2}{c}{1400$^\circ$C} \\
& \multicolumn{2}{c}{Grain Size (\textmu m)} & \multicolumn{2}{c}{Grain Size (\textmu m)} \\
\hline
Manual Segmentation  & \multicolumn{2}{c}{0.41 $\pm$ 0.05} & \multicolumn{2}{c}{2.04 $\pm$ 0.50} \\
Workflow      & \multicolumn{2}{c}{0.54 $\pm$ 0.18} & \multicolumn{2}{c}{1.99 $\pm$ 0.57} \\
NIST GSAT MLI & \multicolumn{2}{c}{0.07 $\pm$ 0.01} & \multicolumn{2}{c}{0.15 $\pm$ 0.03} \\
\hline
\multicolumn{5}{c}{\textbf{Pore Space}} \\
\hline
& \multicolumn{2}{c}{1200$^\circ$C} & \multicolumn{2}{c}{1400$^\circ$C} \\
& Porosity & Mean Pore Dia. & Porosity & Mean Pore Dia. \\
& (\%) & (\textmu m) & (\%) & (\textmu m) \\
\hline
Manual Segmentation & 9.4 $\pm$ 2.0 & & 0.27 $\pm$ 0.20 & \\
Workflow     & 12.4 $\pm$ 2.7  & 0.148 $\pm$ 0.032 & 2.44 $\pm$ 0.68   & 0.074 $\pm$ 0.008 \\
PoreSpy      & 9.8 $\pm$ 1.7   & 0.031 $\pm$ 0.004 & 0.49 $\pm$ 0.22   & 0.019 $\pm$ 0.008 \\
\hline
\end{tabular}
\end{table}



\section{Conclusion}

An automated image analysis pipeline was developed and validated for quantitative characterization of sintered ceramic microstructures. The key outcome of this study was the establishment of a workflow that is in agreement with expert manual segmentation results. This work represents the combination of materials science methodologies with computer science techniques by validating the synthetic data generation, which creates opportunities for data-driven design of materials.
Future work will include the creation of conditional generative models for generating microstructures from a set of specified process parameters, building machine learning frameworks that integrate physical principles, expanding the application to other material systems, and the integration of tomographic data for three-dimensional characterization of sintered ceramics.

\begin{credits}
\subsubsection{\ackname}
This preprint has not undergone peer review (when applicable) or any post-submission improvements or corrections. The Version of Record of this contribution is published in Paszynski, M., Barnard, A.S., Zhang, Y.J. (eds) Computational Science – ICCS 2026 Workshops. Lecture Notes in Computer Science, vol 16787. Springer, Cham, and is available online at \url{https://doi.org/10.1007/978-3-032-29909-3_47}.

We thank Nick Satterlee, John Kang, Elisa Torresani, and Eugene Olevsky for fruitful discussions.
This work was partially supported by the National Science Foundation 
under NSF Awards DMR-2119833 and DMR-2414458. Any opinions, findings and conclusions or recommendations expressed in this material are those of the authors and do not necessarily reflect those of the National Science Foundation. 
This research was supported in part through the use of DARWIN computing system: DARWIN – A Resource for Computational and Data-intensive Research at the University of Delaware and in the Delaware Region, Rudolf Eigenmann, Benjamin E. Bagozzi, Arthi Jayaraman, William Totten, and Cathy H. Wu, University of Delaware, 2021, \url{https://udspace.udel.edu/handle/19716/29071}.
\subsubsection{\discintname}
The authors have no competing interests.
\end{credits}

%
\hbadness=8000
\bibliographystyle{splncs04}
\bibliography{paper}

@misc{Patel2022,
    key = {Patel2022},
    author    = {Patel, A.V. and Hou, T. and Rodriguez, J.D.B. and Dey, T.K. and Birnie, D.P.},
	title = {Patel, {A}.{V}., {Hou}, {T}., {Rodriguez}, {J}.{D}.{B}., {Dey}, {T}.{K}., {Birnie}, {D}.{P}.: {Topological} filtering for {3D} microstructure segmentation. {Computational} {Materials} {Science} 202, 110920 (2022) - {Google} {Search}},
    doi = {10.48550/arXiv.2104.13430},
	urldate = {2026-02-19},
}

@article{Sauvola2000,
	title = {Adaptive document image binarization},
	volume = {33},
	issn = {0031-3203},
	doi = {10.1016/S0031-3203(99)00055-2},
	number = {2},
	urldate = {2026-02-19},
	journal = {Pattern Recognition},
	author = {Sauvola, J. and Pietikäinen, M.},
	month = feb,
	year = {2000},
	pages = {225--236},
}

@book{Bradski2000,
    key = {Bradski2000},
	title = {The {Opencv} {Library}},
	volume = {25},
	author = {Bradski, Gary},
	month = nov,
	year = {2000},
    publisher = {Dr. Dobb's Journal of Software Tools},
	note = {Journal Abbreviation: Dr. Dobb's J. Softw. Tools Publication Title: Dr. Dobb's J. Softw. Tools},
}

@article{VanDerWalt2014,
	title = {scikit-image: {Image} processing in {Python}},
	volume = {2},
	issn = {2167-8359},
	shorttitle = {scikit-image},
	doi = {10.7717/peerj.453},
	urldate = {2026-02-19},
	journal = {PeerJ},
	author = {Walt, Stefan van der and Schönberger, Johannes L. and Nunez-Iglesias, Juan and Boulogne, François and Warner, Joshua D. and Yager, Neil and Gouillart, Emmanuelle and Yu, Tony and contributors, the scikit-image},
	month = jun,
	year = {2014},
	note = {arXiv:1407.6245 [cs]},
	pages = {e453},
}

@inproceedings{Tomasi1998,
	title = {Bilateral filtering for gray and color images},
	doi = {10.1109/ICCV.1998.710815},
	urldate = {2026-02-19},
	booktitle = {Sixth {International} {Conference} on {Computer} {Vision}},
	author = {Tomasi, C. and Manduchi, R.},
	month = jan,
	year = {1998},
	pages = {839--846},
}

@misc{Zuiderveld1994,	
	key = {Zuiderveld1994},
	title = {Contrast limited adaptive histogram equalization {\textbar} {Graphics} gems {IV}},
	doi = {10.5555/180895.180940},
	language = {en},
	urldate = {2026-02-19},
	journal = {Guide books},
}

@article{Vincent1993,
	title = {Morphological grayscale reconstruction in image analysis: applications and efficient algorithms},
	volume = {2},
	issn = {1941-0042},
	shorttitle = {Morphological grayscale reconstruction in image analysis},
	doi = {10.1109/83.217222},
	number = {2},
	urldate = {2026-02-19},
	journal = {IEEE Transactions on Image Processing},
	author = {Vincent, L.},
	month = apr,
	year = {1993},
	pages = {176--201},
}

@misc{Soille2003,
	key = {Soille2003},
	title = {Morphological {Image} {Analysis}: {Principles} and {Applications} {\textbar} {Guide} books {\textbar} {ACM} {Digital} {Library}},
	shorttitle = {Morphological {Image} {Analysis}},
    doi = {10.5555/773286},
	language = {en},
	urldate = {2026-02-19},
	journal = {Guide books},
}

@misc{ASTM2013,
    key = {ASTM2013},
	title = {{ASTM} {E112}-13 (2013) {Standard} {Test} {Methods} for {Determining} {Average} {Grain} {Size}, {ASTM} {International}, {West} {Conshohocken}. - {References} - {Scientific} {Research} {Publishing}},
	url = {https://www.scirp.org/reference/referencespapers?referenceid=3247763},
	urldate = {2026-02-19},
}

@article{Chen2023,
	title = {Additive {Manufacturing} of {Ceramics}: {Advances}, {Challenges}, and {Outlook}},
	volume = {43},
	shorttitle = {Additive {Manufacturing} of {Ceramics}},
	doi = {10.1016/j.jeurceramsoc.2023.07.033},
	journal = {Journal of the European Ceramic Society},
	author = {Dadkhah, Mehran and Tulliani, Jean-Marc and Saboori, Abdollah and Iuliano, Luca},
	month = jul,
	year = {2023},
}

@article{Schmidt2019,
	title = {Recent advances and applications of machine learning in solid-state materials science},
	volume = {5},
	copyright = {2019 The Author(s)},
	issn = {2057-3960},
	doi = {10.1038/s41524-019-0221-0},
	language = {en},
	number = {1},
	urldate = {2026-02-19},
	journal = {npj Computational Materials},
	author = {Schmidt, Jonathan and Marques, Mário R. G. and Botti, Silvana and Marques, Miguel A. L.},
	month = aug,
	year = {2019},
	note = {Publisher: Nature Publishing Group},
	pages = {83},
}

@article{Agrawal2019,
	title = {Deep materials informatics: {Applications} of deep learning in materials science},
	volume = {9},
	issn = {2159-6867},
	shorttitle = {Deep materials informatics},
	doi = {10.1557/mrc.2019.73},
	language = {en},
	number = {3},
	urldate = {2026-02-19},
	journal = {MRS Communications},
	author = {Agrawal, Ankit and Choudhary, Alok},
	month = sep,
	year = {2019},
	pages = {779--792},
}

@article{DeCost2019,
	title = {High {Throughput} {Quantitative} {Metallography} for {Complex} {Microstructures} {Using} {Deep} {Learning}: {A} {Case} {Study} in {Ultrahigh} {Carbon} {Steel}},
	volume = {25},
	copyright = {https://www.cambridge.org/core/terms},
	issn = {1431-9276, 1435-8115},
	shorttitle = {High {Throughput} {Quantitative} {Metallography} for {Complex} {Microstructures} {Using} {Deep} {Learning}},
	doi = {10.1017/S1431927618015635},
	language = {en},
	number = {1},
	urldate = {2026-02-19},
	journal = {Microscopy and Microanalysis},
	author = {DeCost, Brian L. and Lei, Bo and Francis, Toby and Holm, Elizabeth A.},
	month = feb,
	year = {2019},
	pages = {21--29},
}

@article{Chowdhury2016,
	title = {Image driven machine learning methods for microstructure recognition},
	volume = {123},
	issn = {0927-0256},
	doi = {10.1016/j.commatsci.2016.05.034},
	urldate = {2026-02-19},
	journal = {Computational Materials Science},
	author = {Chowdhury, Aritra and Kautz, Elizabeth and Yener, Bülent and Lewis, Daniel},
	month = oct,
	year = {2016},
	pages = {176--187},
}

@article{Sezgin2004,
	title = {Survey over image thresholding techniques and quantitative performance evaluation},
	volume = {13},
	doi = {10.1117/1.1631315},
	journal = {Journal of Electronic Imaging},
	author = {Sezgin, Mehmet and Sankur, Bulent},
	month = jan,
	year = {2004},
	pages = {146--168},
}

@article{Otsu1979,
	title = {A {Threshold} {Selection} {Method} from {Gray}-{Level} {Histograms}},
	volume = {9},
	issn = {2168-2909},
	doi = {10.1109/TSMC.1979.4310076},
	number = {1},
	urldate = {2026-02-19},
	journal = {IEEE Transactions on Systems, Man, and Cybernetics},
	author = {Otsu, Nobuyuki},
	month = jan,
	year = {1979},
	pages = {62--66},
}

@article{Kapur1985,
	title = {A new method for gray-level picture thresholding using the entropy of the histogram},
	volume = {29},
	issn = {0734-189X},
	doi = {10.1016/0734-189X(85)90125-2},
	number = {3},
	urldate = {2026-02-19},
	journal = {Computer Vision, Graphics, and Image Processing},
	author = {Kapur, J. N. and Sahoo, P. K. and Wong, A. K. C.},
	month = mar,
	year = {1985},
	pages = {273--285},
}

@book{Niblack1986,
	title = {An introduction to digital image processing},
	isbn = {978-0-13-480674-7},
	language = {eng},
	urldate = {2026-02-19},
	publisher = {Englewood Cliffs, N.J. : Prentice-Hall International},
	author = {Niblack, Wayne},
	collaborator = {{Internet Archive}},
	year = {1986},
}

@article{Choi2001,
	title = {Automatic grain boundary detection and grain size analysis using polarization micrographs or orientation images},
	volume = {22},
	doi = {10.1016/S0191-8141(00)00014-6},
	journal = {Journal of Structural Geology},
	author = {Heilbronner, Renée},
	month = jul,
	year = {2000},
	pages = {969--981},
}

@article{Decenciere2001,
	title = {Morphological decomposition of the surface topography of an internal combustion engine cylinder to characterize wear},
	volume = {249},
	doi = {10.1016/S0043-1648(01)00579-8},
	journal = {Wear},
	author = {Decencière, Etienne and Jeulin, Dominique},
	month = jun,
	year = {2001},
	pages = {482--488},
}

@article{Gostick2017,
	title = {Versatile and efficient pore network extraction method using marker-based watershed segmentation},
	volume = {96},
	doi = {10.1103/PhysRevE.96.023307},
	number = {2},
	urldate = {2026-02-19},
	journal = {Physical Review E},
	author = {Gostick, Jeff T.},
	month = aug,
	year = {2017},
	note = {Publisher: American Physical Society},
	pages = {023307},
}

@article{Meyer1990,
	title = {Morphological segmentation},
	volume = {1},
	issn = {1047-3203},
	doi = {10.1016/1047-3203(90)90014-M},
	number = {1},
	urldate = {2026-02-19},
	journal = {Journal of Visual Communication and Image Representation},
	author = {Meyer, F. and Beucher, S.},
	month = sep,
	year = {1990},
	pages = {21--46},
}

@article{Adams1994,
	title = {Seeded region growing},
	volume = {16},
	issn = {1939-3539},
	doi = {10.1109/34.295913},
	number = {6},
	urldate = {2026-02-19},
	journal = {IEEE Transactions on Pattern Analysis and Machine Intelligence},
	author = {Adams, R. and Bischof, L.},
	month = jun,
	year = {1994},
	pages = {641--647},
}

@misc{Ronneberger2015,
	title = {U-{Net}: {Convolutional} {Networks} for {Biomedical} {Image} {Segmentation}},
	shorttitle = {U-{Net}},
	doi = {10.48550/arXiv.1505.04597},
	urldate = {2026-02-19},
	publisher = {arXiv},
	author = {Ronneberger, Olaf and Fischer, Philipp and Brox, Thomas},
	month = may,
	year = {2015},
	note = {arXiv:1505.04597 [cs]},
}

@article{Azimi2018,
	title = {Advanced {Steel} {Microstructural} {Classification} by {Deep} {Learning} {Methods}},
	volume = {8},
	copyright = {2018 The Author(s)},
	issn = {2045-2322},
	doi = {10.1038/s41598-018-20037-5},
	language = {en},
	number = {1},
	urldate = {2026-02-19},
	journal = {Scientific Reports},
	author = {Azimi, Seyed Majid and Britz, Dominik and Engstler, Michael and Fritz, Mario and Mücklich, Frank},
	month = feb,
	year = {2018},
	note = {Publisher: Nature Publishing Group},
	pages = {2128},
}

@inproceedings{Buades2005,
	title = {A non-local algorithm for image denoising},
	volume = {2},
	doi = {10.1109/CVPR.2005.38},
	urldate = {2026-02-19},
	booktitle = {2005 {IEEE} {Computer} {Society} {Conference} on {Computer} {Vision} and {Pattern} {Recognition} ({CVPR}'05)},
	author = {Buades, A. and Coll, B. and Morel, J.-M.},
	month = jun,
	year = {2005},
	note = {ISSN: 1063-6919},
	pages = {60--65 vol. 2},
}

@article{Huang1979,
	title = {A fast two-dimensional median filtering algorithm},
	volume = {27},
	issn = {0096-3518},
	doi = {10.1109/TASSP.1979.1163188},
	number = {1},
	urldate = {2026-02-19},
	journal = {IEEE Transactions on Acoustics, Speech, and Signal Processing},
	author = {Huang, T. and Yang, G. and Tang, G.},
	month = feb,
	year = {1979},
	pages = {13--18},
}

@article{Rudin1992,
	title = {Nonlinear total variation based noise removal algorithms},
	volume = {60},
	issn = {0167-2789},
	doi = {10.1016/0167-2789(92)90242-F},
	number = {1},
	urldate = {2026-02-19},
	journal = {Physica D: Nonlinear Phenomena},
	author = {Rudin, Leonid I. and Osher, Stanley and Fatemi, Emad},
	month = nov,
	year = {1992},
	pages = {259--268},
}

@misc{Bernsen1986,
	key = {Bernsen1986},
	title = {J. {Bernsen}, “{Dynamic} {Thresholding} of {Gray} {Level} {Image},” {ICPR}`86 {Proceedings} of {International} {Conference} on {Pattern} {Recognition}, {Berlin}, 1986, pp. 1251-1255. - {References} - {Scientific} {Research} {Publishing}},
	urldate = {2026-02-19},
}

@article{Jaccard1912,
	title = {The {Distribution} of the {Flora} in the {Alpine} {Zone}.},
	volume = {11},
	issn = {1469-8137},
	doi = {10.1111/j.1469-8137.1912.tb05611.x},
	language = {en},
	number = {2},
	urldate = {2026-02-19},
	journal = {New Phytologist},
	author = {Jaccard, Paul},
	year = {1912},
	note = {\_eprint: https://nph.onlinelibrary.wiley.com/doi/pdf/10.1111/j.1469-8137.1912.tb05611.x},
	pages = {37--50},
}

@inproceedings{Rahman2016,
    key = {Rahman2016},
	title = {Optimizing {Intersection}-{Over}-{Union} in {Deep} {Neural} {Networks} for {Image} {Segmentation}},
    booktitle = {Proceedings of the International Symposium on Visual Computing},
	volume = {10072},
	isbn = {978-3-319-50834-4},
	doi = {10.1007/978-3-319-50835-1_22},
	author = {Rahman, Md. Atiqur and Wang, Yang},
	month = dec,
	year = {2016},
	pages = {234--244},
}

@inproceedings{Csurka2013,
	address = {Bristol},
	title = {What is a good evaluation measure for semantic segmentation?},
	isbn = {978-1-901725-49-0},
	doi = {10.5244/C.27.32},
	language = {en},
	urldate = {2026-02-19},
	booktitle = {Procedings of the {British} {Machine} {Vision} {Conference} 2013},
	publisher = {British Machine Vision Association},
	author = {Csurka, Gabriela and Larlus, Diane and Perronnin, Florent},
	year = {2013},
	pages = {32.1--32.11},
}

@misc{Gonzalez2018,
    key = {Gonzalez2018},
	title = {Gonzalez, {R}.{C}. and {Woods}, {R}.{E}. (2018) {Digital} {Image} {Processing}. 4th {Edition}, {Pearson} {Education}, {London}. - {References} - {Scientific} {Research} {Publishing}},
	urldate = {2026-02-19},
}

@article{Munch2008,
	title = {Contradicting {Geometrical} {Concepts} in {Pore} {Size} {Analysis} {Attained} with {Electron} {Microscopy} and {Mercury} {Intrusion}},
	volume = {91},
	copyright = {© 2008 The American Ceramic Society},
	issn = {1551-2916},
	doi = {10.1111/j.1551-2916.2008.02736.x},
	language = {en},
	number = {12},
	urldate = {2026-02-19},
	journal = {Journal of the American Ceramic Society},
	author = {Münch, Beat and Holzer, Lorenz},
	year = {2008},
	note = {\_eprint: https://ceramics.onlinelibrary.wiley.com/doi/pdf/10.1111/j.1551-2916.2008.02736.x},
	pages = {4059--4067},
}

@article{Abrams1971,
	title = {Grain size measurement by the intercept method},
	volume = {4},
	issn = {0026-0800},
	doi = {10.1016/0026-0800(71)90005-X},
	number = {1},
	urldate = {2026-02-19},
	journal = {Metallography},
	author = {Abrams, Halle},
	month = feb,
	year = {1971},
	pages = {59--78},
}

@article{Lopez-Sanchez2020,
	title = {Which average, how many grains, and how to estimate robust confidence intervals in unimodal grain size populations},
	volume = {135},
	issn = {0191-8141},
	doi = {10.1016/j.jsg.2020.104042},
	urldate = {2026-02-19},
	journal = {Journal of Structural Geology},
	author = {Lopez-Sanchez, Marco A.},
	month = jun,
	year = {2020},
	pages = {104042},
}

@incollection{Goldstein2018,
	address = {New York, NY},
	title = {{SEM} {Image} {Interpretation}},
	isbn = {978-1-4939-6676-9},
	language = {en},
	urldate = {2026-02-19},
	booktitle = {Scanning {Electron} {Microscopy} and {X}-{Ray} {Microanalysis}},
	publisher = {Springer},
	author = {Goldstein, Joseph I. and Newbury, Dale E. and Michael, Joseph R. and Ritchie, Nicholas W. M. and Scott, John Henry J. and Joy, David C.},
	editor = {Goldstein, Joseph I. and Newbury, Dale E. and Michael, Joseph R. and Ritchie, Nicholas W.M. and Scott, John Henry J. and Joy, David C.},
	year = {2018},
	doi = {10.1007/978-1-4939-6676-9_7},
	pages = {111--121},
}

@article{Gostick2019,
	title = {{PoreSpy}: {A} {Python} {Toolkit} for {Quantitative} {Analysis} of {Porous} {Media} {Images}},
	volume = {4},
	shorttitle = {{PoreSpy}},
	doi = {10.21105/joss.01296},
	journal = {Journal of Open Source Software},
	author = {Gostick, Jeff and Khan, Zohaib Atiq and Tranter, Thomas and Kok, Matthew and AGNAOU, Mehrez and Sadeghi, Amin and Jervis, Rhodri},
	month = may,
	year = {2019},
	pages = {1296},
}

@misc{Moser2022,
	key = {Moser2022},
	title = {usnistgov/grain-size-analysis-tools},
	url = {https://github.com/usnistgov/grain-size-analysis-tools},
	urldate = {2026-02-19},
	publisher = {National Institute of Standards and Technology},
	month = feb,
	year = {2022},
	note = {original-date: 2024-07-10T21:47:29Z},
}

@article{Schroeder2020,
	title = {The {ImageJ} ecosystem: {Open}‐source software for image visualization, processing, and analysis},
	volume = {30},
	issn = {0961-8368},
	shorttitle = {The {ImageJ} ecosystem},
	doi = {10.1002/pro.3993},
	number = {1},
	urldate = {2026-02-20},
	journal = {Protein Science : A Publication of the Protein Society},
	author = {Schroeder, Alexandra B. and Dobson, Ellen T. A. and Rueden, Curtis T. and Tomancak, Pavel and Jug, Florian and Eliceiri, Kevin W.},
	month = jan,
	year = {2021},
	pmid = {33166005},
	pmcid = {PMC7737784},
	pages = {234--249},
}

@article{Girard2011,
	title = {{LOCAL} {STUDY} {OF} {DEFECTS} {DURING} {SINTERING} {OF} {UO2}: {IMAGE} {PROCESSING} {AND} {QUANTITATIVE} {ANALYSIS} {TOOLS}},
	volume = {27},
	copyright = {Copyright (c) 2014 Image Analysis \& Stereology},
	issn = {1854-5165},
	shorttitle = {{LOCAL} {STUDY} {OF} {DEFECTS} {DURING} {SINTERING} {OF} {UO2}},
	doi = {10.5566/ias.v27.p79-85},
	language = {en},
	number = {2},
	urldate = {2026-02-20},
	journal = {Image Analysis and Stereology},
	author = {Girard, Eric and Chaix, Jean-Marc and Valdivieso, François and Goeuriot, Patrice and Lechelle, Jacques},
	year = {2008},
	pages = {79--85},
}

@inproceedings{hadjadj2016isauvola,
  author    = {Hadjadj, Zineb and Meziane, Abdelkrim and Cherfa, Yazid 
               and Cheriet, Mohamed and Setitra, Insaf},
  title     = {{ISauvola}: Improved {Sauvola's} Algorithm for Document 
               Image Binarization},
  booktitle = {Image Analysis and Recognition. ICIAR 2016},
  series    = {Lecture Notes in Computer Science},
  volume    = {9730},
  pages     = {737--745},
  publisher = {Springer, Cham},
  year      = {2016},
  doi       = {10.1007/978-3-319-41501-7_82}
}

\end{document}